\pdfoutput=1
\documentclass[conference,compsoc]{IEEEtran}
\usepackage{tikz}
\usepackage{amsmath}
\usepackage{ifthen}
\usepackage{comment}
\excludecomment{confidential}
\usepackage[symbol]{footmisc}
\usepackage{xspace}
\usepackage{amsmath}
\usepackage{amssymb}
\usepackage{amsthm}
\usepackage{mathpartir}
\usepackage{mathtools}
 \usepackage{stmaryrd}
\usepackage{calc}
\usepackage{pifont}
\usepackage{graphicx}
\graphicspath{{./figures/}}
\DeclareGraphicsExtensions{.pdf,.png}
\usepackage{threeparttable}
\usepackage{listings}
\usepackage{paralist}
\usepackage{enumitem}
\usepackage[nocompress]{cite}
\usepackage{hyperref}
\usepackage{cleveref}
\usepackage{balance}
\usepackage{url}
\usepackage{color}
\usepackage{colortbl}
\usepackage{xcolor}
\usepackage{framed}
\colorlet{shadecolor}{yellow!10}
\newcommand*\colourcheck[1]{
      \expandafter\newcommand\csname #1check\endcsname{\textcolor{#1}{\ding{52}}}
  }
\colourcheck{red}
\colourcheck{green}
\definecolor{ao(english)}{rgb}{0.0, 0.5, 0.0}
\definecolor{royalblue(web)}{rgb}{0.25, 0.41, 0.88}
\newcommand{\code}[1]{{\small\texttt{#1}}}

\newcommand{\term}[1]{{\em{#1}}}
\newcommand{\myparagraph}[1]{\smallskip \noindent{\textbf{{#1}.~}}}
\newcommand{\myparagraphnodot}[1]{\smallskip \noindent{\textbf{{#1}}}}
\newcommand{\secref}[1]{$\S$\ref{sec:#1}}
\newcommand{\subsecref}[1]{$\S$\ref{subsec:#1}}

\newcommand{\appref}[1]{Appendix~\ref{app:#1}}

\newcommand{\figref}[1]{Figure~\ref{fig:#1}}
\newcommand{\tabref}[1]{Table~\ref{tab:#1}}

\newcommand{\defref}[1]{Definition~\ref{def:#1}}

\newcommand{\program}{p}
\newcommand{\attack}{\mathit{Measure}}
\newcommand{\contract}{\mathit{Contract}}

\newcommand{\cnt}{\mathit{obs}}

\theoremstyle{definition}
\newtheorem{definition}{Definition}

\renewcommand{\program}{p}
\newcommand{\archstate}{\sigma}
\newcommand{\microstate}{\mu}

\definecolor{Blue3}{HTML}{0000CD}

\makeatletter
\renewcommand{\xRightarrow}[2][]{\ext@arrow 0359\Rightarrowfill@{#1}{#2}}
\makeatother
\mathchardef\hyphenmathcode=\mathcode`\-
\lstdefinestyle{embedded}{
    basicstyle=\ttfamily\small,
    numbers=left,
    numberstyle=\tiny,
    numbersep=3pt,
    numberblanklines=true,
    frame=tb,
    aboveskip=5pt,
    belowskip=5pt,
    columns=fullflexible,
    showstringspaces=false,
    keepspaces=true,
    showlines=true,
    xleftmargin=8pt,
    backgroundcolor=\color{gray!4},
    framesep=1pt,
}
\lstdefinestyle{asm}{
    language={[x86masm]Assembler},
    basicstyle=\footnotesize\ttfamily,
    numbers=left,
    numberstyle=\tiny,
    numbersep=3pt,
    numberblanklines=true,
    frame=tb,
    aboveskip=0pt,
    belowskip=0pt,
    columns=fullflexible,
    showstringspaces=false,
    keepspaces=true,
    showlines=true,
    xleftmargin=8pt,
    backgroundcolor=\color{gray!4},
    morekeywords={MFENCE,R14,RAX,RBX,RCX,RDX,CMOVNL,CMOVNZ},
    morecomment=[l]{\#},
    commentstyle=\itshape\color{gray},
    keywordstyle=\ttfamily,
    tabsize=4,
    keywordstyle=\color{blue!80!black},
    identifierstyle=\color{green!20!black},
}

\lstdefinestyle{contract}{
    basicstyle=\footnotesize\ttfamily,
    numbers=none,
    aboveskip=0pt,
    belowskip=0pt,
    columns=fullflexible,
    showstringspaces=false,
    keepspaces=true,
    showlines=true,
    xleftmargin=8pt,
    backgroundcolor=\color{gray!4},
    morekeywords={ld, str, pc},
    morecomment=[l]{\#},
    commentstyle=\itshape\color{gray},
    keywordstyle=\ttfamily,
    tabsize=4,
    keywordstyle=\color{blue!80!black},
    identifierstyle=\color{green!20!black},
}

\lstdefinelanguage{yaml}{
    morecomment=[l]{\#},
    commentstyle=\color{gray}
}

\let\origlstlisting=\lstlisting
\let\endoriglstlisting=\endlstlisting
\renewenvironment{lstlisting}
{\mathcode`\-=\hyphenmathcode
\everymath{}\mathsurround=0pt\origlstlisting}
{\endoriglstlisting}

\newboolean{publicversion}
\setboolean{publicversion}{true}
\ifthenelse{\boolean{publicversion}}{
\newcommand{\grumbler}[2]{}
\newcommand{\grumblerb}[2]{}
\newcommand{\grumblerp}[2]{}
\newcommand{\grumblerc}[2]{}
\newcommand{\editing}[1]{}

}{
\newcommand{\grumbler}[2]{\textcolor{red}{{{\bf #1}: #2}}}
\newcommand{\grumblerb}[2]{\textcolor{blue}{{{\bf #1}: #2}}}
\newcommand{\grumblerp}[2]{\textcolor{purple}{{{\bf #1}: #2}}}
\newcommand{\grumblerc}[2]{\textcolor{cyan}{{{\bf #1}: #2}}}
\newcommand{\grumblerg}[2]{\textcolor{green}{{{\bf #1}: #2}}}
\newcommand{\grumblerm}[2]{\textcolor{magenta}{{{\bf #1}: #2}}}

\newcommand{\editing}[1]{{\color{Sepia}#1}}

}

\iftrue
\newcommand{\fs}[1]{\grumblerc{\scriptsize FS}{\scriptsize  \bf#1}}
\newcommand{\oo}[1]{\grumblerp{\scriptsize OO}{\scriptsize \bf#1}}
\newcommand{\jh}[1]{\grumblerb{\scriptsize JH}{\scriptsize \bf#1}}
\newcommand{\bk}[1]{\grumbler{\scriptsize BK}{\scriptsize \bf#1}}
\newcommand{\cf}[1]{\grumblerg{\scriptsize CF}{\scriptsize \bf#1}}
\newcommand{\sv}[1]{\grumblerm{\scriptsize SV}{\scriptsize \bf#1}}

\newcommand{\todo}[1]{ \editing{TODO: #1}}
\else
\newcommand{\fs}[1]{}
\newcommand{\oo}[1]{}
\newcommand{\jh}[1]{}
\newcommand{\bk}[1]{}
\newcommand{\cf}[1]{}
\newcommand{\sv}[1]{}

\newcommand{\todo}[1]{}
\fi

\usepackage{tcolorbox}
\usepackage{xcolor}
\usepackage{colortbl}

\newif\ifsubmission
\submissionfalse

\newcounter{ChallengeCounter}
\begin{document}

\title{Enter, Exit, Page Fault, Leak\IEEEauthorrefmark{3}:\\
    Testing Isolation Boundaries for Microarchitectural Leaks}

\author{\IEEEauthorblockN{Oleksii Oleksenko\IEEEauthorrefmark{1}}
\IEEEauthorblockA{Azure Research\\
Microsoft}
\and
\IEEEauthorblockN{Flavien Solt\IEEEauthorrefmark{1}}
\IEEEauthorblockA{ETH Zurich\IEEEauthorrefmark{2}}
\and
\IEEEauthorblockN{Cédric Fournet}
\IEEEauthorblockA{Azure Research\\
Microsoft}
\and
\IEEEauthorblockN{Jana Hofmann}
\IEEEauthorblockA{MPI-SP\IEEEauthorrefmark{2}}
\and
\IEEEauthorblockN{Boris Köpf}
\IEEEauthorblockA{Azure Research\\
Microsoft}
\and
\IEEEauthorblockN{Stavros Volos}
\IEEEauthorblockA{Azure Research\\
Microsoft}}

\maketitle

\footnotetext[1]{Joint first authors.}
\footnotetext[2]{Work partially done while at Azure Research, Microsoft.}
\footnotetext[3]{The title references ``Tinker, Tailor, Soldier, Spy'', a novel about an MI6 agent uncovering a mole that was leaking secrets to the Soviet Union.}

\begin{abstract}

    CPUs provide isolation mechanisms like virtualization and privilege levels to protect software. 
    Yet these focus on architectural isolation while typically overlooking microarchitectural side channels, exemplified by Meltdown and Foreshadow. 
    Software must therefore supplement architectural defenses with ad-hoc microarchitectural patches, which are constantly evolving as new attacks emerge and defenses are proposed.
    Such reactive approach makes ensuring complete isolation a daunting task, and leaves room for errors and oversights.

    We address this problem by developing a tool that \emph{stress tests microarchitectural isolation} between security domains such as virtual machines, kernel, and processes, with the goal of detecting flaws in the isolation boundaries.
    The tool extends model-based relational testing (MRT) methodology to enable detection of cross-domain information leakage.
    We design a new test case generator and execution sandbox to handle multi-domain execution, new leakage models to encode expected leaks, and new analysis techniques to manage nondeterminism.

    We use this tool to perform an in-depth testing campaign on six x86-64 CPUs for leakage across different isolation boundaries.
    The testing campaign exposed four new leaks and corroborated numerous known ones, with only two false positives throughout the entire campaign. 
    These results show critical gaps in current isolation mechanisms as well as validate a robust methodology for detecting microarchitectural flaws.
    As such, this approach enables a shift from reactive patching to proactive security validation in processor design.

\end{abstract}

\section{Introduction}
\label{sec:intro}

Modern systems rely on \emph{security domain} abstractions, such as OS kernels with user processes and hypervisors with virtual machines (VMs).
Though they run on the same hardware, these domains must be logically isolated:
a domain must neither observe nor influence the state of equal- or higher-privilege domains, communicating exclusively through well-defined interfaces.

Numerous architectural mechanisms are in place to enforce this isolation: page tables prevent one process from accessing another's memory; virtualization conceals hypervisor data from VMs; privilege levels bar user processes from system configuration.
Yet these mechanisms often fall short of full isolation, as shared microarchitectural state (e.g., caches or CPU buffers) remains vulnerable to attacks like Meltdown~\cite{Kocher2018,Lipp2018} or Foreshadow~\cite{Bulck2018}.
Architectural mechanisms thus must be complemented with microarchitectural isolation mechanisms to prevent information leaks through shared hardware.

The rollout of such microarchitectural protections, however, is driven by the urgency of vulnerability response, and it results in a fragmented landscape:
some leaks are mitigated in hardware or firmware, others in the OS or hypervisor (e.g. flushing of buffers), and yet some others are not patched at all because their security impact is considered low and the mitigation cost is high.

This fragmentation complicates efforts to seal microarchitectural leaks between security domains.
The current practice is to either confirm that proof-of-concept attacks stop working, or to use purely theoretical reasoning~\cite{Canella2020}, which is laborious and error-prone.
Consequently, there have been several cases of incomplete or even completely ineffective patches, and vendors' claims about microarchitectural security are routinely disproven~\cite{bhi, inception,inspectreGadget,cacheout,milburn2022win,wikner2022retbleed,weber2023reviving,Hofmann23}.

The situation calls for a change.
There needs to be a way to systematically test microarchitectural isolation mechanisms, in their various forms and with varying levels of opaqueness.
Yet the existing solutions are limited:
prior work provides only point solutions, targeting a single domain (e.g., kernel or user space), a single kind of leakage, or a single mitigation~\cite{Nemati2020a,oleksenko2022revizor,weber2023reviving,Moghimi2020a}.
We aim to fill this gap and offer a comprehensive alternative.

\myparagraph{Approach}
In this paper, we present the first tool to systematically test microarchitectural isolation between various security domains.
Our design relies on the methodology of model-based relational testing (MRT)~\cite{oleksenko2022revizor,Nemati2020a,oleksenko2023hide,Hofmann23}, an emerging approach that has been successful in automatically detecting a broad spectrum of microarchitectural leaks.

MRT is a random testing technique: the key idea is to execute random programs on the target CPU, record their leakage, and compare it against an abstract model of the known leaks. If the observed leakage does not match the model prediction, it means that the CPU leaks unexpected information, and this is reported to the user.

Unfortunately, existing MRT approaches operate strictly within a single domain, namely in kernel mode.
 Leveraging MRT to analyze leakage across domain transitions requires tackling the following challenges.
\begin{compactenum}
\item We need to {\em generate test cases} that randomly explore many different microarchitectural conditions, but that also execute the specific instruction sequences required for transitioning between domains and for specific mitigations.
There are many domain transitions, mitigations, and configuration options, and we want a uniform way to handle them all.
\item We need to {\em execute these test cases and capture the information they leak}. This requires (a) setting up the environments for executing code in different domains, and for transitioning between them, (b) at a speed that facilitates exploring a large number of test cases. We furthermore need to (c) deal with the nondeterminism introduced by domain transitions and the nontrivial amounts of opaque microcode they trigger.
\item We need to specify {\em leakage models} for domain transitions. Such models need to reflect which domains are considered adversarial and which need to be protected.
\end{compactenum}

We solve these challenges by introducing an {\em actor} framework. 
We use ``actor'' as an umbrella abstraction for representing any security domain, which encompasses processes, VMs, or other isolated execution contexts.
On a technical level, an actor is a region of code and data that is executed with a specific privilege level, CPU mode, and with specific memory permissions; and that is presumably isolated from other actors by hardware primitives.
One actor can transition to another one by executing a specific sequence of instructions (e.g., syscall or VM enter).

We design a {\em test case generator} that allows us to create code that represents different patterns of interacting actors.
It takes as input a template of actors and transitions between them (e.g., OS entering a user process) and it instantiates the template with random code as well as pre-defined instruction sequences for managing the transitions and (optionally) implementing any leakage mitigations under test.
We develop an {\em execution environment} for running multi-actor test cases and for gathering measurement traces.
The execution environment configures the actor's context by setting up address spaces, creating VMs, etc.
It relies on live OS modification to provide test cases with safe and full access to the underlying hardware within an unmodified operating system, and without introducing crashes.
We define {\em leakage models} for actor-based test cases. The baseline model is to require that no information about the victim's state is leaked, but we also accommodate weaker forms of isolation by supporting leakage models that permit leakage of memory access patterns.

Finally, we develop novel techniques for {\em robust measurement} of hardware traces to cope with nondeterminism. We do this by lifting tests of equality of traces to tests of equality of the distributions from which they are drawn, which enables us to leverage established statistical tools.

\myparagraph{Experiments}
Our techniques are built as an extension to Revizor~\cite{oleksenko2022revizor}, an open-source tool for detecting microarchitectural vulnerabilities using MRT. We demonstrate their effectiveness by testing isolation between different security domains on six modern CPUs (three Intel and three AMD).
Our experiments cover four patterns of domain transitions, each tested in over twenty attack scenarios, totalling billions of separate hardware measurements and 88 machine-days of testing.
We highlight the following results:
\begin{asparaenum}
    \item Our tool discovered \emph{four novel} cross-domain leaks.
    The first exposes cached memory from one VM to another;
    the second exposes recently-stored values from kernel to user mode;
    and the other two show that AMD CPUs speculate on certain privileged instruction executed in user mode.
    \item The tool detected \emph{all known types} of cross-domain leaks that are possible to surface in our experimental setup, including MDS and its variants~\cite{RIDL,ZombieLoad}, Foreshadow~\cite{Bulck2018,weisse2018foreshadow}, Meltdown~\cite{Lipp2018}, DSS~\cite{Hofmann23}, and Meltdown 3a~\cite{RSRR}.
    \item In terms of \emph{performance}, our tool performs 800--4500 measurements per second, yielding a throughput of 60--700 test case executions per second.
    \item Our robust analysis techniques successfully handled measurement nondeterminism, with only \emph{two false positives} during the entire campaign, enabling us to focus human investigations on interesting test cases.
\end{asparaenum}

\myparagraph{Summary of Contributions}
We develop techniques and a tool that enable comprehensive automatic testing of microarchitectural isolation between security domains.
We use our tool to perform a broad testing campaign on commercial x86 CPUs where we surface several instances of unknown leaks, demonstrating the effectiveness and practicality of our approach.

\myparagraph{Responsible Disclosure}
We disclosed our discoveries to the CPU vendors in June 2024.
Following the reports, AMD issued CVE-2024-36357, CVE-2024-36350, CVE-2024-36349, and CVE-2024-36348.
The corresponding security bulletin detailing the security impact and the mitigation information is available at \url{https://www.amd.com/en/resources/product-security/bulletin/amd-sb-7029.html}

\myparagraph{Availability}
The described tooling is available as a part of Revizor v1.3+:
\url{https://github.com/microsoft/sca-fuzzer}

\section{Background}
\label{sec:bak}

We provide an overview of the main isolation abstractions on x86, and the hardware primitives that are used to enforce microarchitectural isolation between them.
We then give an overview of Model-based Relational Testing, a technique for detecting microarchitectural information leaks.

\subsection{Isolation Abstractions}
Computer systems provide various levels of accesses to resources, allowing them to run multiple security domains while ensuring hierarchical resource 
isolation. Our work focuses on security domains that are enabled by (1) virtualization, which allows hypervisors to manage and virtualize hardware resources to implement 
virtual machines; and (2) protection rings, which encapsulate hierarchical levels of privilege 
and allow operating systems (OS) to manage resources to implement \textit{processes}.

\myparagraph{Virtual Machine (VM) Isolation} 
VMs are the foundational block of today's clouds as they enable providers to share large compute systems
across tenants while providing strong isolation. A VM emulates an entire computer system, and runs its 
own operating system, isolated from other VMs. Extended (Intel) or Nested (AMD) Page Tables (EPTs/NPTs) enforce memory isolation between VMs.
Each VM possesses its own E/NPT, managed by the hypervisor, which maps guest physical memory to host physical memory.

The per-VM control structure (called VMCS on Intel, and VMCB on AMD) is a hypervisor-managed data structure that configures VM execution.
Notably, the VMCS/B defines conditions that trigger a \texttt{VMEXIT}, for example certain guest instructions can be configured to trap to the hypervisor. 
On the contrary, the execution context changes from host (hypervisor) to guest (VM) through a \texttt{VMLAUNCH} (Intel) or \texttt{VMRUN} (AMD) instruction.

\myparagraph{User-Kernel Isolation}
User programs run within user processes, which are assigned resources by the OS kernel.
Each process owns its own virtual address space which is mapped to (host or guest) physical memory space based on the process's page table. Page tables are managed 
by the OS kernel in such way that its resources are not accessible by user processes, and processes are isolated from each other. 

A user process transitions to kernel either explicitly via system calls or implicitly via exceptions.
The transfer back is triggered by the \code{SYSRET} instruction.

\subsection{Microarchitectural Isolation Primitives on x86}

Security researchers and CPU vendors are constantly working on defenses to combat microarchitectural vulnerabilities.
Examples include flushing microarchitectural store buffers with the legacy \texttt{VERW} instruction to mitigate some MDS attacks~\cite{Fallout}, 
flushing the L1-D cache on context switches, kernel page table isolation (KPTI)~\cite{LinuxKPTI}, divider state flushing~\cite{Hofmann23}, and 
indirect branch prediction barriers (IBPB) to mitigate branch prediction sharing~\cite{yavarzadeh2024pathfinder}.
Additionally, Intel and AMD have implemented hardware patches to address vulnerabilities like Meltdown, Foreshadow, and MDS.
However, there is a constant stream of new vulnerabilities and mitigations have been found to be (partially) ineffective~\cite{bhi, inception,inspectreGadget,cacheout,milburn2022win,wikner2022retbleed}.

\subsection{Model-Based Relational Testing and Revizor}
\label{subsec:mrt}

In our work, we use Model-based Relational Testing (MRT)~\cite{oleksenko2022revizor} as a methodology to search for unexpected information leakage in CPUs.
MRT relies on a model that encodes \emph{known} microarchitectural vulnerabilities to predict the information leaked when executing some code, which allows it to distinguish between expected and unexpected leaks.
The approach generates random code that is executed both in the CPU-under-test and in the model. It then measures the microarchitectural state changes caused by the code and compares it to the leakage predicted by the model.
If the observed leakage matches the model's prediction, this indicates that the CPU is behaving as expected, and the test case is discarded.
Otherwise, if the random code exposed \emph{unexpected} information, this indicates a potential security vulnerability, and the generated code can be used as a starting point for further (manual) analysis of the new leak.
The methodology has been shown to be effective in finding both new vulnerabilities and variants of known vulnerabilities~\cite{oleksenko2022revizor,oleksenko2023hide,Hofmann23,Nemati2020a}.

In our implementation, we rely on one specific tool that implements the MRT methodology: Revizor~\cite{oleksenko2022revizor}.
It works by executing the following loop for a number of rounds, or until a model violation is detected:

\begin{asparaenum}
    \item \emph{Program Generation}: A testing round starts by generating a program.
    The program is essentially a random sequence of assembly instructions generated from a predefined instruction pool.
    The generator can be configured to constrain the shape of the program's control-flow graph, control the pool of instructions, and configure the instruction frequencies.
    It also (optionally) instruments the program to prevent undesired faults, such as division by zero.
    \item \emph{Input Generation}: The next step is to generate a set of program inputs.
    Each input is a file with binary contents, used to initialize the program's memory and registers.
    The input generator populates the files with values from a (seeded) pseudo-random number generator.
    A combination of a program and a sequence of inputs is called a \emph{test case}.
    \item \emph{Hardware Execution}: The executor takes a program, executes it on the target CPU with each of the inputs, and collects the observable microarchitectural state changes for each execution.
    Such changes are called \emph{hardware traces}.
    They are typically collected via a side-channel attack, such as Prime+Probe, in which case a trace is a set of cache lines evicted by the program.
    In addition to collecting traces, the executor also ensures a low-noise and reproducible execution environment, for example, by disabling interrupts and flushing caches before starting a measurement.
    \item \emph{Model Execution}: Similarly, the model takes the program and executes it with each of the inputs.
    The model is based on an ISA emulator (namely, Unicorn~\cite{quynh2015unicorn}), modified to record the data that we expect to be leaked on the given CPU, and to emulate the expected speculative behavior.
    For example, a model may record the addresses of all loads and stores (i.e., the information learned via a cache side channel), and emulate branch prediction by temporarily taking a wrong branch target on every conditional jump.
    This way, a model \emph{overapproximates} the information that we expect to be observable by an attacker.
    \item \emph{Comparing Leakage}: 
    The trace analyzer uses the leakage predicted by the model to filter out instances of expected leaks from the set of collected hardware traces, thus leaving only the unexpected leaks.
    The instances where hardware traces expose information that is not predicted by the model are reported to the user.
    The analyzer performs the filtering by checking the noninterference property~\cite{Guarnieri2021} w.r.t. the model;
    we further elaborate on how this approach checks isolation in~\secref{model}.
\end{asparaenum}

The key aspect of this approach is that MRT never \emph{directly} compares the hardware traces to the model's prediction; instead, it compares the exposed \emph{information}.
This allows a complex modern CPU to be tested against a simple model, and still effectively filter out expected leaks while detecting unexpected leaks.

\section{Overview}
\label{sec:overview}

In this work, we test isolation boundaries between security domains.
To do so, we design and implement an \emph{actor}-based framework that builds on Revizor, a state-of-the-art open-source tool for detecting microarchitectural leaks with model-based relational testing (MRT).

Actors represent distinct and generally mutually distrusting security domains that constitute a test case; a test case consists of one 'main' actor executed in the kernel host mode, and one or more actors in other modes. 
Each actor comprises of a region of code executed in a given CPU mode and privilege level\footnote{Only host-kernel, host-user, and guest-kernel are currently supported.}, and a region of private data memory
with configurable permissions and other properties. Each actor may have multiple entry and exit points; an exit from one actor is a transition to a different actor, 
except for the 'main' actor that may terminate the test case.

This paper describes modifications to Revizor to enable the following:
\begin{asparaenum}[(1)]
    \item generate multi-actor test cases (\secref{testcases});
    \item execute them in a configurable environment that supports different types of actors and that can collect their hardware traces (\secref{executor});
    \item execute multi-actor test cases in a model describing the expected level of isolation between actors (\secref{model}); 
    \item compare hardware and contract traces while tolerating nondeterminism introduced by the execution of opaque microcode involved in domain transitions (\secref{noise}).
\end{asparaenum}

\myparagraph{Test Case Generation} 
Our test case generator creates code that represents different patterns of interacting actors.   
Unlike the original version of Revizor, which operated on a single domain and relied on fully random contexts (program and input), 
testing domain interactions requires testing of a pre-defined sequence of instructions in a randomized context. 

Our test case generator takes as input a template of actors and transitions between them (e.g., Figure~\ref{fig:asm_template}), 
instantiates the template with pre-defined code for managing these transitions, and populates each 
actor with random code based on a pool of instructions specified in a configuration file, such as the one shown in Figure~\ref{fig:config_file}. 

 \myparagraph{Execution Environment} 
We execute test cases in a highly configurable execution environment (e.g., CPU configuration, 
actor's context, page tables) without sacrificing performance required for MRT to be effective. 

To this end, we rely on a custom Linux kernel module (executor), 
which takes control over the host and configures the context of actors as prescribed in the template and the configuration file. 
The executor also implements low-overhead routines for management tasks such as creating virtual machines, page tables, and configuring system registers, and implements facilities for collecting hardware traces (e.g., Prime+Probe routines).

\myparagraph{Leakage Models}
As we are interested in information leakage between security domains, our models need to describe leakage at the level of actors rather than at the level of instructions. 
We therefore model multi-actor execution and specify a suitable version of the noninterference property (\S\ref{sec:model}). 
We use Revizor's infrastructure to execute these models within the Unicorn emulator~\cite{quynh2015unicorn}, which we extend to domain transitions (\S\ref{sec:loading}).

\section{Test Cases Generation}
\label{sec:testcases}

In this section, we describe how the generator creates multi-actor test cases.

\subsection{Requirements}

To test domain transitions and the effectiveness of microarchitectural isolation primitives, we need to test them in various contexts. As such, our test case generator needs to fulfill two requirements:
\begin{asparaitem}
    \item \emph{Structuring}: A specific sequence of instructions is required to perform a domain transition and/or apply a microarchitectural isolation primitive.  
    For example, (1) to transition from a hypervisor to a virtual machine (VM), the hypervisor actor must execute a specific sequence of instructions to set up the VM exit and entry points, and then execute a 
    VM enter instruction; and (2) many microarchitectural isolation primitives are implemented as sequences of instructions (e.g., L1D cache flush) that have to be executed right before a domain transition.
    As such, the test cases need to be structured in a way that the framework can generate code that involves a \emph{pre-defined} instruction sequence and a \emph{randomized} context.        
    \item \emph{Unification:} Test cases need to be generated for two different stages of the testing pipeline: executor and model. However, some of the \emph{pre-defined} instruction sequences need to be different
    across these two stages due to incompleteness of the ISA supported by the model's emulator. For example, in a hypervisor-VM transition, the hypervisor actor in the emulator needs to involve a different instruction sequence for than the one involved by the hypervisor actor 
    in the executor due to lack of emulation of the VM enter/resume instructions. As such, the test cases need to defined in a unified way while enabling the framework to instantiate the same test case in different ways according to the stage of 
    the testing pipeline.
\end{asparaitem}

\subsection{Defining Test Case Structure}

We structure the test cases while providing a degree of randomness by introducing two concepts: templates and macros. 
\term{Template} is an assembly file that defines (1) the structure of a test case by combining valid assembly code with macros and (2) the boundaries of each actor's code by placing them in different sections.
\term{Macro} is a special pseudo-instruction that looks like a no-op in the assembly code but can be instantiated with a complex operation by different stages of the testing pipeline, as discussed in \S\ref{sec:loading}.

\figref{asm_template} illustrates these two concepts through an example of two interacting actors, \code{main} (lines~1--10) and \code{guest} (lines~12--17).
The configuration file in~\figref{config_file} specifies that \code{main} runs in host mode with kernel privileges, while \code{guest} runs in guest mode and also kernel privileges.
The template starts by entering the \code{main} actor, followed by a sequence of 64 randomly generated instructions (line~3 in~\figref{asm_template}), a flush of memory buffers with the \code{VERW} instruction (line~4), and a transition to the \code{guest} actor (lines~5--7).
After the transition, the \code{guest} actor starts measurements (line~14), executes another sequence of randomly generated instructions (line~15), ends measurements (line~16), and transitions back to the \code{main} actor (line~17), which completes the test case.
Should any of the random instructions trigger an exception, the template defines the location of the fault handler at line~9.

Note that we intentionally use a standard assembly syntax for implementing templates and macros, as it means that templates are valid assembly files, and it allows us to re-use standard tooling such as GNU assembler and ELF parser.

\begin{figure}[t]
    \begin{center}
        \begin{lstlisting}[style=asm]
.section .main
.start:
  .macro.random_instructions.64:
  VERW qword ptr [r14]
  .macro.set_h2g_target.vm_start:  # set VM entry
  .macro.set_g2h_target.end:       # set VM exit
  .macro.switch_h2g:
.end:
.macro.fault_handler:
# test case exits upon reaching this location

.section .guest
.vm_start:
  .macro.measurement_start:
  .macro.random_instructions.64:
  .macro.measurement_end:
  .macro.switch_g2h:
\end{lstlisting}
    \end{center}
    \caption{Example of a test case template that defines two actors---main and guest---and the transitions between them.}
    \label{fig:asm_template}
\end{figure}

\subsection{Generation Process}
\label{subsec:generation}

Our test case generator takes a template and a configuration file (written by the user) as input and creates test case programs.
If the template includes at least one instance of the \code{random\_instructions} macro, the generator can create many test cases from a single template, allowing us to test the same transition pattern in different contexts.

\myparagraph{Program Generation}
The first step is to expand the \code{random\_instructions} macros in the template.
The generator takes the template, parses it into an internal representation, replaces each instance of the macro with a sequence of random instructions, and instruments the instructions to prevent undesired exceptions/faults.
The number of instructions is determined by the first argument of the macro (line~3 of \figref{asm_template}).
The pool of instructions is defined in the configuration file (line~11 in \figref{config_file}).
Once the assembly program is generated, it is assembled into an ELF binary.

\begin{figure}[t]
  \begin{center}
      \begin{lstlisting}[style=asm,language=yaml]
actors:
- main:                      
  - mode: "host"
  - privilege_level: "kernel"             
- guest:                        
  - mode: "guest"
  - privilege_level: "kernel"            
  - observer: true           
  - data_properties:         
    - writable: false             
instruction_allowlist:
- ... # omitted for brevity
contract_observation_clause: load+store+pc
contract_execution_clause:
- noninterference
enable_prefetchers: false\end{lstlisting}
  \end{center}
  \caption{Example of a configuration file complementing the template in \figref{asm_template}.
    The file specifies two actors, their modes, privilege levels, and data properties.
      The \texttt{instruction\_allowlist} property lists the instructions that could be used to generate test cases based on the template.
      The guest is an observer, meaning that it attempts to leak information from the main actor.
      The contract specifies that the observer actor (guest) cannot observe any information belonging to the main actor, as described in~\secref{model}.}
  \label{fig:config_file}
\end{figure}

Initially, each actor's randomized context includes memory accesses to the actor's private memory. However, it is useful to include a subset of 
memory accesses that target global memory or private memory of other actors. For instance, we may want to test (1) whether private memory of one actor 
is leaked to another actor during explicit communication between actors via global memory, or (2) scenarios for specific microarchitectural leaks 
that arise due to one actor accessing memory that is only present in the address space of another actor. 

To support such scenarios, we introduce a 
new instrumentation pass, which randomly selects one of the generated memory accesses and modifies its address such that it points to a different actor's memory. 
In our current implementation, this pass specifically targets Meltdown leaks by moving user-mode actor's memory accesses to the memory of the \code{main} actor, 
which is always a kernel-mode actor. (As shared memory is out of scope (\S\ref{sec:discussion}), we leave a general implementation of this instrumentation pass for future work.)

\myparagraph{Input Generation}
Together with a generated program, we also generate a sequence of random inputs to it. 
Specifically, for each actor, we generate three pages of random data to initialize its data memory and registers.

\begin{figure}[t]
  \begin{center}
      \includegraphics[width=1.\columnwidth]{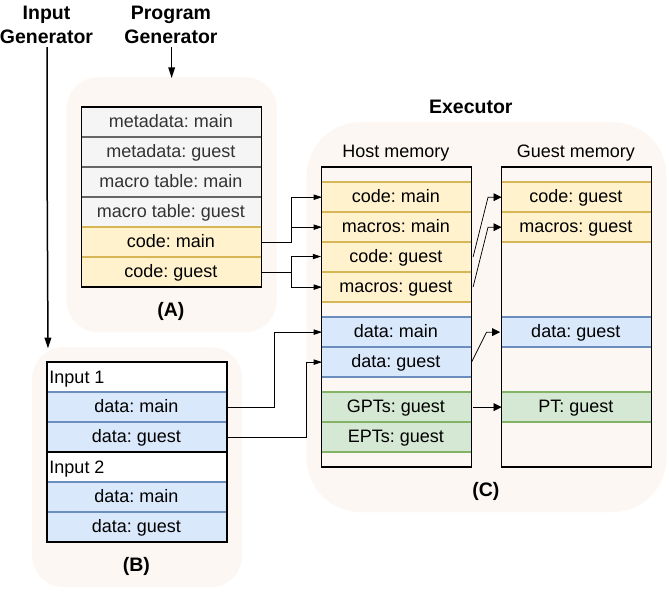}
  \end{center}
  \caption{Loading of a test case into executor.
  (a) The binary test case produced by the program generator.
  (b) The binary with a sequence of inputs.
  (c) The executor memory after loading the test case.}
  \label{fig:transfer}
\end{figure}

\myparagraph{Assembling Test Cases}
The next step is to combine the test case binary and its configuration into a package that can be loaded into the executor and the model.
Note that we could not rely solely on the ELF binary as it does not include information about macros and actor configurations.
Instead, we create a custom binary format as shown in~\figref{transfer}a.

The generator first populates the metadata for each actor with the information from the configuration file (\figref{transfer}a).
This includes the actor's mode, privilege level, and data permissions.
Then it creates a table describing macros, in a format similar to a symbol table in an ELF binary.
The table lists all macros in the program, along with their owner actors, offset in the binary, and arguments.
Finally, the generator copies the compiled binary of every actor from the ELF file into the corresponding code sections.

Similarly, the input generator copies the generated inputs into the corresponding sections of the binary package, as shown in~\figref{transfer}b.

\subsection{Implementation of Macros}
\label{sec:loading}
The macros in the binary package (\figref{transfer}a) are merely placeholder NOPs, and it is the responsibility of the executor and the model to interpret them correctly.

\myparagraphnodot{In the executor}, macros are implemented via binary patching.
The executor first allocates dedicated code memory for every actor (e.g., \figref{transfer}c, ``code: main''), and then copies each actor's code into the corresponding memory region.
After that, for each macro in the macro tables (\figref{transfer}a), the executor inserts an implementation of this macro into a dedicated memory region (e.g., ``macros: main'' in \figref{transfer}c), then replaces the NOP instruction in the actor's code with a jump to the corresponding macro implementation, and places a jump back to the original location at the end of the macro implementation.
If a macro is supposed to receive arguments, then the executor hardcodes them into the macro implementation.
This way, a macro is replaced with a pseudo-call to a routine that implements the macro's behavior.

Note that we rely on direct jumps for the implementation, which are the least intrusive way to implement control-flow changes in x86 CPUs, which is important for keeping the microarchitectural measurements precise.
The only known type of speculation that may affect jumps is straight-line speculation~\cite{SLS}, which we prevent by inserting an \code{LFENCE} after the macro jumps.

The specific routine that implements a macro is determined by the target CPU and the configuration file.
For example, the \code{switch\_h2g} macro, which transitions from the host to the guest actor, is replaced by a jump to a routine that executes \code{VMRESUME} on Intel CPUs, but on AMD CPUs it would be replaced by a jump to a routine that executes \code{VMRUN}.
Similarly, the \code{measurement\_start} macro, which starts collecting hardware traces, is replaced by a jump to a routine that implements Prime if the requested measurement mode is Prime+Probe, or to a routine that implements Flush if the requested mode is Flush+Reload.

\myparagraphnodot{In the model}, macros are implemented as dynamic callbacks.
The model has a hook function invoked before every instruction, which we extend to track macros.
Whenever the model encounters a NOP with an address matching a macro in the macro table, it calls the corresponding macro implementation.
For example, when the model encounters a NOP at the address of the \code{switch\_h2g} macro, it calls a function that emulates a switch to VM mode and redirects the emulator to the entry point of the guest actor.
\section{Execution Environment}
\label{sec:executor}

This section discusses requirements and design of the execution environment for multi-domain test cases. 
We focus on the delta to prior work~\cite{oleksenko2022revizor,Hofmann23}, which already provides an environment for single-domain (kernel) execution.

\subsection{Requirements}
To execute a multi-domain test case, we have to set up a rich execution environment---create page tables, prepare virtual machines for VM actors, set initial state, etc---and then execute the test case and collect hardware traces.
This setup code has to fulfill two requirements:
\begin{asparaitem}
    \item \emph{Full Configurability}:
          To cater for any domain interaction pattern, we need to be able to freely configure all aspects of the execution environment, from page table permissions, to the CPU mode, and to VM configurations.
          This also implies permitting invalid system configurations, as those might be necessary to trigger certain leakage types.
    \item \emph{Low Overhead}:
          The effectiveness of MRT techniques depends on the number of test cases that can be executed per unit of time.
          To be on par with the current SoA in MRT-based tools~\cite{oleksenko2022revizor,oleksenko2023hide,Hofmann23}, we aim for at least a hundred measurements per second, which is why we need ability to set up VMs and create page tables in milliseconds.
\end{asparaitem}

\vspace{1mm}

\subsection{Design and Implementation}

Because of these requirements, we could not reuse the existing implementation of the system management features in Linux:
First, the implementation of VM and page table management in Linux is targeted towards full-scale VMs, and hence, it introduces too much overhead for tiny VM actors that we test.
Second, the Linux kernel puts a significant number of guard rails in place to ensure stability and security of the system, which would prevent us from freely configuring the experiments.
As a result, we decided to implement a custom execution environment from scratch, totaling at over 3'500 lines of code in a kernel module.

The complete setup and measurement process is shown in~\figref{measurement_algorithm}.
Below we describe the key ideas behind it.

\begin{figure}[t]
    \begin{center}
        \begin{lstlisting}[style=asm,language=C]
traces = [];
disable_interrupts();
preserve_host_state();
configure_system_registers();
create_virtual_machines();
create_page_tables();
flush_caches_and_buffers();

for (input in inputs)
    set_actor_data_memory(input);
    set_page_table_permissions();
    set_interrupt_descriptor_table();
    set_general_purpose_registers(input);
    htrace = execute(test_case_program);
    traces.append(htrace);
    restore_page_table_permissions();
    restore_interrupt_descriptor_table();

restore_host_state();
enable_interrupts();
return traces;
\end{lstlisting}
    \end{center}
    \caption{High-level algorithm of setting up the execution environment for multi-actor test cases.}
    \label{fig:measurement_algorithm}
\end{figure}

\myparagraph{State Preservation and Fault Isolation}
As multi-domain measurements involve reconfiguration of the CPU on the fly, we need to carefully manage the CPU state so as to avoid corrupting the host Linux kernel and crashing the system.
To this end, the executor stores the host OS state at the beginning of the experiment (line~3 in~\figref{measurement_algorithm}) and restores it at the end (line~18).
The preserved state includes control registers (e.g., \code{CR0}), model-specific registers (e.g., \code{EFER}), virtualization state (e.g., control registers of VMX on Intel CPUs), and the state of host page tables entries that we modify during the experiments.

In addition, we must take care of the faults that might occur during the execution of the test case.
Our tool permits actors to trigger arbitrary faults and exceptions, as those might be necessary to trigger certain leakage types.
To ensure that the host OS is not affected by these faults, we catch all exceptions and direct them to a custom fault handler (line~13 and 17 in~\figref{measurement_algorithm}).
Simple exceptions (e.g., page faults) are directed either to the test case exit point or, if \code{fault\_handler} macro is present, to its location, thus allowing the test case to handle exceptions internally.
Low-level interrupts (e.g., NMI or machine check) are directed to custom handlers that recover the original OS state and then exit the executor.
We also disable all interrupts on the core to prevent unexpected exits (line~2).

\myparagraph{CPU Configuration}
To achieve a high level of control over the execution environment, the executor provides a rich configuration interface to set up the CPU under test.
The interface allows to modify performance counters, MSRs, and other system registers (line~4 in~\figref{measurement_algorithm}).
This permits the executor to enable/disable CPU features as well as support VM and user actors (described below).

A user of our tool can control these features through a configuration file.
For example, line~16 of~\figref{config_file} requests that prefetchers should be disabled during the measurements.
The executor implements it by selecting the appropriate MSR for the CPU model under test, and sets the corresponding bits to disable the prefetchers.
Notably, the tool currently supports a only small subset of all possible configuration options, but we implement the interface in such a way that new options can be added trivially.

\myparagraph{VM Actor Configuration}
If the test case includes at least one VM actor, the executor performs additional steps to create the VMs.
First, it enables the virtualization extensions, either Intel VMX or AMD SVM (line~4 in~\figref{measurement_algorithm}), and initializes the VM control structures (VMCS or VMCB) for each guest actor (line~5 in~\figref{measurement_algorithm}).
Second, the executor creates the nested/extended page tables and guest page tables for each guest actor (line~6 in~\figref{measurement_algorithm}), thus providing each VM with its own private memory space.
Finally, it sets permissions on the guest and nested/extended page tables according to actor configuration (line~11 in~\figref{measurement_algorithm}).
Note that both VM configuration and page table permissions can be later modified by the test case itself using macros.

For example, the configuration file in~\figref{config_file} specifies one VM actor (\code{guest}) that has a read-only data page (lines~5--10).
Assuming that the target CPU is an Intel CPU, the executor creates an environment for this actor by creating a VMCS for it, creating a set of guest and extended page tables as shown in~\figref{transfer}c, and clearing the ``writable'' bit in the guest page table entry for the actor's data page.

\myparagraph{User Actor Configuration}
If the test case includes at least one user actor, the executor updates the system enter/exit points to point to the user actor's code (line~4 in~\figref{measurement_algorithm}), and updates page table permissions for the actor's pages to allow user access.
Similar to VM actors, the enter/exit points can be later modified by the test case using macros.

\myparagraph{Memory Aliasing}
We also implement an optional feature that allows to maximize the chances of speculative address confusions, similar to the Foreshadow-VM attack~\cite{weisse2018foreshadow}.
When this feature is enabled, the executor creates an identical memory layout for all VM actors, that is, all code and data pages of all VM actors will be located at the same virtual addresses.
On top of it, the guest physical addresses will match the host physical addresses of the \code{main} actor.
\section{Modeling of Domain Isolation}
\label{sec:model}

In this section, we present the isolation property underlying our methodology and the model we employ to efficiently find cross-domain information leakage.

\subsection{Multi-Actor Noninterference}
The traditional property describing isolation between several users is noninterference~\cite{goguen1982security}.
We use a variant of noninterference, which we adapt to the multi-actor setting.
To simplify the description, we assume the setting with two actors, a victim (e.g., hypervisor) and an attacker (e.g., VM).
Noninterference states that the victim's (potentially secret) data must not influence the measurements conducted by the attacker (i.e., the hardware traces).
We formulate it as follows:
Let $\sigma = (\sigma^V, \sigma^A)$ be the test case data (i.e., the contents of the memory), which is composed of the victim's data $\sigma^V$ and the attacker's data $\sigma^A$ (see \figref{transfer}c).
The attacker's measurements are the outputs of a function $\attack{}(\program{},\archstate{},\microstate{})$, 
defined on data $\sigma$, an initial microarchitectural state $\mu$, and a program $p$ which encompasses the instructions executed by the attacker and the victim.
The victim is isolated from the attacker if $\attack{}(\program{},\archstate{},\microstate{})$ is independent of $\sigma^V$:
\begin{align*}
    \sigma^A_1 = \sigma^A_2 \Rightarrow \attack{}(\program{},\archstate{}_1,\microstate{}) = \attack{}(\program{},\archstate{}_2,\microstate{})
\end{align*}

\subsection{Contract-Based Noninterference}

In practice, most operating systems and hypervisors do not provide such strong isolation guarantees but permit some basic leakage.
For example, the return-from-syscall routine in the Linux kernel \emph{does not} flush caches, meaning that the user process is implicitly permitted to observe the recent kernel's load/store addresses and jump/call targets.
It \emph{does} execute memory buffer resets to prevent MDS-like attacks, so any other parts of the kernel data should not leak.

We therefore weaken the noninterference property by allowing some leakage on the victim's side, namely the leakage of victim's load/store addresses and call/jump targets.

To model this permitted leakage, we rely on the prior work on leakage contracts~\cite{Guarnieri2018}.
We use an emulator-based model (see~\subsecref{mrt}) that collects addresses and control-flow changes to produce a \emph{contract trace}.
We modify the model to support multiple actors, and to collect traces only for the victim actor.
We use the resulting contract trace to permit leakage of victim's addresses of memory accesses and control-flow changes.
Abstractly, the contract trace is defined as a function $\contract{}(\program{},\archstate{}^V)$, that depends on the program $\program{}$ and the victim's data $\sigma^V$.
This results in the following isolation property.
\begin{definition}\label{def:isolation}

    For a two-actor setting with a victim $V$ and an attacker $A$, the CPU satisfies isolation if for any program $p$, any two architectural states $\sigma_1$ and $\sigma_2$, and any initial microarchitectural state $\mu$, it holds
    \begin{align*}
        &\sigma^A_1 = \sigma^A_2 ~\land~ \contract{}(\program{},\archstate{}^V_1) = \contract{}(\program{},\archstate{}^V_2) \\
        &\Rightarrow \attack{}(\program{},\archstate{}_1,\microstate{}) = \attack{}(\program{},\archstate{}_2,\microstate{})
    \end{align*}
\end{definition}
With the above definition, isolation is only violated if there is \emph{more} leakage than the accessed memory addresses and control flow changes.

Note that Definition~\ref{def:isolation} permits more expressive contracts that allow for speculation on the victim's side~\cite{Guarnieri2021,Hofmann23}. We discuss this generalization in~\secref{discussion}.

\myparagraph{Example}
Example contract traces are depicted in~\figref{contract-trace}.
We execute a program on three different inputs $\sigma^V_1, \sigma^V_2$, and $\sigma^V_3$. 
The first two have different values stored in \code{rax}. This implies that the contract traces will be different, meaning that the model predicts that the attacker observes different hardware traces. In this case, we do \emph{not} report a violation, thereby filtering out less interesting cases.
Inputs $\sigma^V_2$, and $\sigma^V_3$, on the other hand, agree on the value of \code{rax}, resulting in identical contract traces.
Now, if the target CPU is vulnerable to MDS, the attacker code in line~12 may speculatively return the value stored at line~8, resulting in different hardware traces if $\sigma^V_2(\texttt{[0xb]}) \neq \sigma^V_3(\texttt{[0xb]})$.
This violates~\defref{isolation}, and the tool flags it as such.

\subsection{Model Implementation}
\label{subsec:model-impl}

We implement the multi-domain model as an extension of the Revizor model, which in turn is based on the Unicorn emulator~\cite{quynh2015unicorn}.
The challenge of extending the model was in the fact that the emulator was designed for single-domain emulation (user mode), and implementing full system emulation would require a significant amount of work.
Instead, we opted for a simpler solution: we executed all actors in user mode, emulated the transitions as simple jumps, and implemented the minimal necessary level of emulation to mirror the behavior of the CPU for a given actor (e.g., a VM exit for privileged instructions).

Note that this approach is sufficient for our purposes, as the MRT methodology permits the model to diverge from the actual CPU behavior, as long as it captures all the information that we expect to leak (e.g., addresses of memory accesses and actor transitions).

\begin{figure}[t]
\begin{minipage}[b]{0.45\linewidth}
\centering
\small
program $p$\\[1em]
\begin{lstlisting}[style=asm]
.section.guest
  # start measuring
  .macro.switch_h2g:

.section.main
  mov rbx, [rax]
  add rbx, 1
  mov [rax], rbx
  .macro.switch_g2h:

.section.guest
  # attacker code
  # end measuring
\end{lstlisting}
\end{minipage}
\begin{minipage}[b]{0.17\linewidth}
\centering
\small
contract\\
trace 1\\[1em]
\begin{lstlisting}[style=contract]


    

pc 0x10
ld 0xa

str 0xa
pc 0x20




\end{lstlisting}
\end{minipage}
\begin{minipage}[b]{0.17\linewidth}
\centering
\small
contract\\
trace 2\\[1em]
\begin{lstlisting}[style=contract]

    
 

pc 0x10
ld 0xb

str 0xb
pc 0x20

 


\end{lstlisting}
\end{minipage}
\begin{minipage}[b]{0.18\linewidth}
\centering
\small
contract\\
trace 3\\[1em]
\begin{lstlisting}[style=contract]

    
  
    
pc 0x10
ld 0xb

str 0xb
pc 0x20
    
  

    
\end{lstlisting}
\end{minipage}
\caption{Simple program and resulting contract traces if $\sigma^V_1(\texttt{rax}) = \texttt{0xa}, \sigma^V_2(\texttt{rax}) = \sigma^V_3(\texttt{rax}) = \texttt{0xb}$,
and the entry point of the host and guest is \texttt{0x10} and \texttt{0x20}, respectively.}
\label{fig:contract-trace}
\end{figure}
\section{Coping with Nondeterminism}
\label{sec:noise}

When testing domain transitions (in particular, VM enters/exits), we encountered a significant increase in the noise, caused by the microcode executed upon a transition.
We could not eliminate it as we do not know exactly what the microcode does, and hence we developed a technique to cope with such nondeterministic measurements.

\myparagraph{The Problem of Nondeterminism}
\defref{isolation} relies on two kinds of equality checks---equality of contract traces and of hardware traces---to determine whether a test case violates the isolation property.
The underlying assumption here is that the difference between two hardware traces is caused by the difference between the inputs to the program.
This is not the case, however, in presence of nondeterminism.
When a pair of hardware traces is affected by noise, they may differ even if there is no genuine information leakage (in fact, two traces may differ even if the inputs are identical).
This leads to frequent false positives, which are time-consuming to investigate and limit the scalability of the tool.

\myparagraph{Lifting Equality to Trace Samples}
To eliminate false positives, we replace an equality test of {\em traces} by an equality test on {\em samples of traces}; that is, we replace direct comparison of values with statistical comparison of distributions.
For choosing a statistical test, we observe that there is no natural order on hardware traces, which is why we treat them as categorical data.
For such data, Pearson's $\chi^2$ test is the standard statistical tool.
For determining whether a sample, assumed independent and identically distributed, of $N$ traces $\overline{t}={t_1,t_2,\dots t_N}$ is drawn from a fixed, predefined distribution $P$, one computes 
\begin{equation*}
\chi^2=\sum_{t\in \overline{t}}(\cnt_{\overline{t}}(t)- N \!\cdot\! P(t))^2/(N \!\cdot\! P(t))
\end{equation*}
Here $\cnt_{\overline{t}}(t)$ describes the number of times $t$ is observed in $\overline{t}$ and $N\!\cdot\! P(t)$ describes how often we would expect to observe $t$ if the sample were drawn from distribution $P$.

The $\chi^2$ test can be leveraged to determine whether two samples $\overline{t_1}$ and $\overline{t_2}$ come from the same (but unknown) distribution. For this, one uses the average count $(\cnt_{\overline{t_1}}(t))+\cnt_{\overline{t_2}}(t))/2$ of each trace in both samples as a proxy for unknown expected count $N\!\cdot\!P$, and computes the average $\chi^2$ value of both samples $\overline{t_2},\overline{t_2}$. If this value is below a chosen threshold,
we conclude the samples $\overline{t_1}$, $\overline{t_2}$ are identically distributed, i.e., the equality check succeeds.

\myparagraph{Performance Optimization: Adaptive Sample Size}
Getting reliable results from statistical tests requires a sufficiently large sample size, meaning that every hardware measurement has to be repeated dozens or even hundreds of times (depending on the noise level).
This is unacceptable for our tool, as it significantly reduces the coverage achievable within a unit of testing time.

To address this issue, we take an iterative approach to choosing sample sizes:
We begin with a small sample of $N=15$ traces per input.
If we detect a violation, we iteratively increase sample sizes to $N=40, 160, 320$\footnote{We found this specific sequence of sizes empirically, while trying to find balance between performance and false negatives in our experiments. They may differ for other experimental setups.}, and report a violation only if the violation persists up to the largest sample size.
This way, we trade off the precision of the tests (i.e., rate of false negatives) for the higher throughput of testing.
In our experiments, this approach reduced the detection time of known leaks by an order of magnitude compared to always using the largest sample size, while still successfully eliminating false positives.

\section{Evaluation}
\label{sec:eval}
In this section, we test isolation boundaries between several security domains on a diverse range of x86 CPUs and for a variety of attack scenarios.
The experiments demonstrate the effectiveness of our tool in:
(1) detecting previously undocumented microarchitectural leakage across isolation boundaries  (\subsecref{new-leaks});
(2) detecting known leaks across isolation boundaries (\subsecref{corroborated});
and (3) testing existing microarchitectural isolation primitives and checking the claims made by CPU vendors about which leaks exist on which CPUs (\subsecref{sw-validation}).

\subsection{Methodology}
Our tests span three dimensions: the isolation boundary, the test target, and the configuration.
Each testing campaign tests 100K random programs generated from a given template, and each program is run with 50 random inputs.
The sample size for each program-input bundle is adaptive starting with 15, and increases iteratively as discussed in~\secref{noise}.
Each combination of template and configuration is seeded with the same seed value across all testing targets to make results comparable.
In each experiment, the executor is configured to use Flush+Reload; i.e., a hardware trace is a set of memory addresses (modulo cache line size) accessed by the test case between \code{measurement\_start} and \code{measurement\_end}.

If a violation is detected, its details are stored and the campaign terminates.
If a patch is available, we apply it and re-run the test to check whether the patch is effective.
We manually inspect every detected violation to ensure that it is a true positive, and to identify the root cause.

\myparagraph{Isolation Boundaries}
We target four types of leakage:
\begin{inparaenum}[(1)]
    \item leakage of host data to VM (\code{H2V});
    \item leakage of data from one VM to another VM (\code{V2V});
    \item leakage of kernel data to user (\code{K2U}); and
    \item leakage of data from one user process to another user process (\code{U2U}).
\end{inparaenum}

To target each type, we build a template with the required actors and domain transitions, and configure their mode and privilege level. For example, the \code{V2V} template contains one host (kernel) actor and two guest (kernel) actors (\code{VM1, VM2}) and the following transitions: \code{Host}$\rightarrow{}$\code{VM2}$\rightarrow{}$\code{Host}$\rightarrow{}$\code{VM1}$\rightarrow{}$\code{Host}$\rightarrow{}$\code{VM2}.
\code{VM2} performs the measurement, and is labeled as an observer (i.e., the attacker).
The code within \code{VM1} and \code{VM2} is randomly generated.

\myparagraph{Testing Targets}
We test the following x86 CPUs, which include both pre- and post-Meltdown models:
\begin{asparaitem}
    \item \code{Int1}: Intel 06-8e-0a (KabyLake R)
    \item \code{Int2}: Intel 06-0e-d (Coffee Lake)
    \item \code{Int3}: Intel 06-b7-1 (RaptorCove)
    \item \code{AMD1}: AMD 17-18-1 (Zen1)
    \item \code{AMD2}: AMD 17-68-1 (Zen2)
    \item \code{AMD3}: AMD 19-74-1 (Zen4)
\end{asparaitem}

\myparagraph{Configurations}
We target three classes of configurations, each representing a scenario in which (at least) one previously known speculative leak could occur.
A configuration describes the pool of instructions that can be used for program generation, the instrumentation passes applied to generated programs, and the permissions on actor's memory pages.
By default, we use the instruction pool\footnote{The default instruction pool is base x86-64 excluding indirect jumps, privileged instructions, and several instructions incorrectly emulated by Unicorn.}
in~\appref{isa-base}, apply instrumentation passes that prevent all faults/exceptions, and permit all memory accesses.
Each configuration modifies the defaults as described below.

\emph{(1) MEM.} The first class centers around memory errors: page faults and memory-related assists.
Each of the configuration generates errors by modifying a bit in the page table permissions of the observer actor (e.g., \code{VM2} in \code{V2V}), as shown in~\tabref{mem-config}.
In all of the configurations, we apply the aliasing memory layout described in~\secref{executor}. In the \code{U-bit} configuration, we also apply the generation pass that inserts random memory accesses from user space to kernel memory described in~\subsecref{generation}.

\begin{table}
    \center
    \footnotesize
    \begin{tabular}{l|l|l}
        \textbf{Type} & \textbf{Error} & \textbf{Description}         \\
        \hline
        \hline
        A-bit         & assist         & Accessed bit is 0            \\
        D-bit         & assist         & Dirty bit is 0               \\
        E/NPT A-bit   & assist         & Accessed bit in EPT/NPT is 0 \\
        E/NPT D-bit   & assist         & Dirty bit in EPT/NPT is 0    \\
        P-bit         & \#PF           & Present bit is 0             \\
        R-bit         & \#PF           & Reserved bit is 1            \\
        W-bit         & \#PF           & Read/Write bit is 0          \\
        E/NPT P-bit   & \#VMEXIT       & Present bit in EPT/NPT is 0  \\
        E/NPT R-bit   & \#VMEXIT       & Reserved bit in EPT/NPT is 1 \\
        E/NPT W-bit   & \#VMEXIT       & Write bit in EPT/NPT is 0    \\
        U-bit         & \#PF           & User bit is 0                \\
        \hline
    \end{tabular}
    \label{tab:mem-config}
    \vspace{1mm}
    \caption{Variants of the \texttt{MEM} configuration.}
\end{table}

\emph{(2) COMP.}
The second class centers around leaks due to speculation upon a computational error, and in particular due to divider state sampling (DSS)~\cite{Hofmann23}.
The errors are triggered by \emph{disabling} the instrumentation pass that otherwise prevents divisions-by-zero by the actor representing the attacker.

\emph{(3) REG.}
The third class centers around leaks from privileged registers to less privileged domains, known as rogue system register reads (RSRR)~\cite{RSRR}.
These leaks are triggered by adding privileged instructions to the instruction pool of the actor representing the attacker, as described in~\appref{isa-extended}.
\subsection{Testing Results}
\label{subsec:testing-results}

Our evaluation consisted of 304 separate testing campaigns that together took 88 machine/days to complete.
These campaigns have tested almost 30 million programs, and have performed over 20 billion hardware measurements.
We found 4 new cross-domain leaks, successfully detected all 6 known leaks possible in the configurations that we tested and reproduced 6 known flaws in patches.
In all of these experiments, we had only two false positives\footnote{The false positives occurred because of higher-than-usual noise bursts, which increased the level of nondeterminism and thus bypassed the statistical test in~\secref{noise}.}, which we discarded after manual inspection.

\begin{figure}[t]
    \begin{center}
        \includegraphics[width=1\columnwidth]{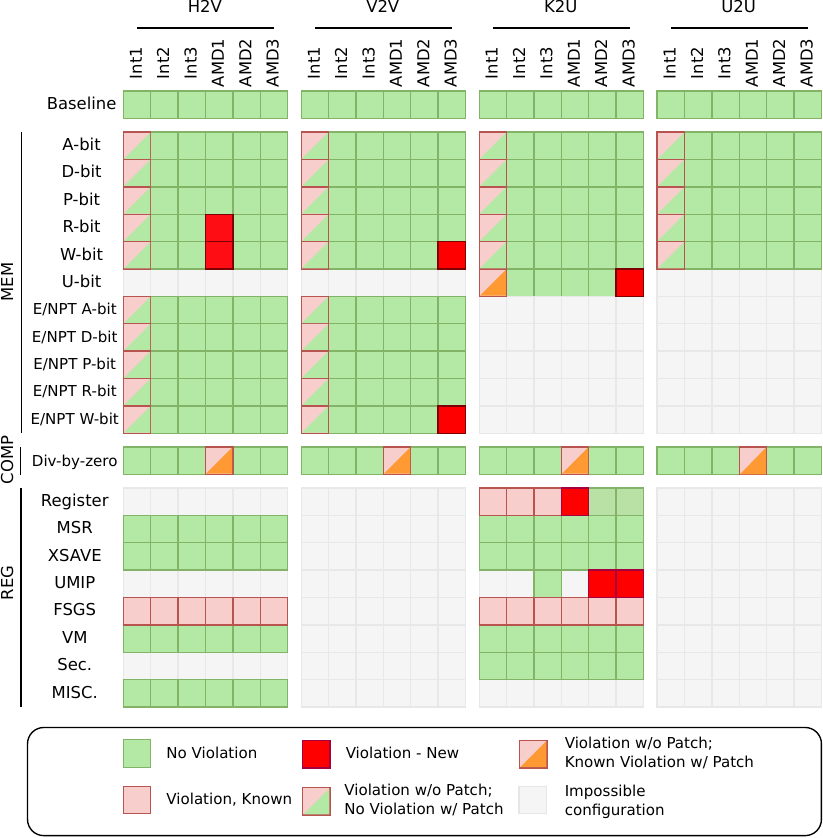}
    \end{center}
    \caption{Overview of the testing results; see text for the label descriptions.}
    \label{fig:eval-overview}
\end{figure}

\figref{eval-overview} provides an overview of the test results.
Due to space constraints, we discuss only selected results.

\subsubsection{New Findings}
\label{subsec:new-leaks}

We found the following new types of cross-domain information leakage.

\myparagraph{Cross-VM Leakage}
We discovered a microarchitectural effect on \code{AMD3} that allows an attacker VM to selectively infer any bit from the memory of another VM, provided the victim VM has previously accessed (i.e., cached) this bit.
By repeatedly exploiting this effect, the attacker VM could read the victim's memory one bit at a time, and thus learn the contents of an arbitrary memory range used by the victim.
The violation was discovered in \code{V2V} template with \code{MEM W-bit} and \code{MEM EPT W-bit} configurations.

Specifically, we observe that
when the attacker VM loads from an address, and the cache contains a value from a victim VM with the {\em same} virtual address (but {\em different} host physical address), the victim's cached value affects the timing of the attacker's operations following the load.
The attacker can test for these timing differences to infer one bit of information about the victim's cached value.
Moreover, the attacker can control the bit to be observed by performing shifts after the load.

The gadget is shown in~\figref{zen4-cross-vm}, and it is executed by the attacker.
It consists of an aliasing load (line~1), followed by two write exceptions (lines~4 and 6), two non-faulting read operations (lines~7 and 8), and finally a memory access using the result of the last read operation (line~10).
Depending on the contents of victim's VM memory, lines~2--9 have different execution times, which means that the attacker can observe the victim's memory by checking whether the last read (line~10) had evicted a cache line.
Effectively, the gadget creates a race between the timing of the exceptions (lines~4 and 6) and the code executed (out-of-order) after the exceptions.

\begin{figure}
    \begin{center}
        \begin{lstlisting}[style=asm, language=C]
a = *p1; // p1 aliases an address in victim VM
array1[a] = 0;
// p2 and p3 point to read-only addresses
*p2 = 0; // page fault
...      // a sequence of arithmetic ops.
*p3 = 0; // page fault
b = *p4; // p4 points to the same page as p2/p3
c = *p5; // p5 points to arbitrary valid address
...      // a sequence of arithmetic ops.
d = array2[c];  // exposure in cache
\end{lstlisting}
    \end{center}
    \caption{Pseudo-code of the gadget that creates cross-VM leak on Zen 4.}
    \label{fig:zen4-cross-vm}
\end{figure}

We would like to note that, at the moment of writing, we do not yet know the root cause of this leak (the details will be disclosed by AMD in the corresponding security advisory).
Hence, it is entirely possible that the leak can be exploited via other---possibly simpler---gadgets, and that the gadget we found is just one of many possible ways to trigger the effect.

This leak has been assigned \code{CVE-2024-36357}.
Its security impact could be potentially high, as it allows an attacker to read arbitrary memory of another VM.

The mitigations for this leak have been developed and deployed during the course of this paper's embargo, and are detailed in the following advisory: \url{https://www.amd.com/en/resources/product-security/bulletin/amd-sb-7029.html}

\myparagraph{Kernel-to-User Leakage}
We discovered a microarchitectural effect on \code{AMD3} that allows a user process to observe any bit
written in any of the last $N$ stores (with $N\approx 32$ in our experiments),
even if the stores were performed in kernel mode.
Using this effect, an unprivileged attacker could extract privileged data from the kernel by scanning its stored values one bit at a time. 
The violation was discovered in \code{K2U} template with \code{MEM U-bit} configuration.

Specifically, we identify a user-level gadget whose timing demonstrates a dependency on data that was stored in kernel mode a fixed number of store operations ago. The attacker can select which of the previous stores to target by performing additional stores. As in the previous leak, the attacker can control which bit of information to leak though timing by performing shifts. 
The attack gadget is similar to the one in Figure~\ref{fig:zen4-cross-vm}, except that it relies on an exception triggered by a user access to kernel memory.

This leak has been assigned \code{CVE-2024-36350}.
Its security impact is potentially moderate-to-high, as the attacker can observe privileged stores preceding switch to the user mode.

Similarly to the previous case, the mitigations are detailed in: \url{https://www.amd.com/en/resources/product-security/bulletin/amd-sb-7029.html}

\myparagraph{Rogue Read of RDTSCP-AUX}
We discovered that a user process can speculatively read the auxiliary timestamp counter register (\code{RDTSCP AUX}), even if such a read is architecturally disabled (\code{TSD} flag is set in \code{CR4}) and thus causes a fault (\code{\#GP}).
The behavior affects \code{Int1, Int2, Int3} and \code{AMD1}.
The violation was discovered in \code{K2U} template with \code{REG-Register} configuration.

This leak has been assigned \code{CVE-2024-36349}.
Its security impact is likely low as the \code{AUX} register usually contains the CPU ID, which is not considered sensitive information.
In fact, this leak was already documented by Intel, but a white paper by AMD claimed that their CPUs are not vulnerable to RSRR~\cite{amdwhitepaper}.
Our results contradict this claim.

Since the leaked information is not considered security sensitive, AMD is not planning to fix this effect.
This is in line with the previous decision by Intel in a similar case.

\myparagraph{Rogue Execution of SMSW}
We discovered that AMD CPUs speculatively execute the \code{SMSW} instruction in user mode, even if the \code{UMIP}~\cite{AMDAPM} feature is enabled and thus architecturally its execution causes a fault (\code{\#GP}).
The speculatively returned value is the 16 lower significant bits of the \code{CR0} register.
The leak affects \code{AMD2} and \code{AMD3} CPUs.
The violation was discovered in \code{K2U} template with \code{REG-UMIP} configuration.

This leak has been assigned \code{CVE-2024-36348}.
Its security impact is likely low as the lower bits of \code{CR0} only contain the CPU configuration.
However, this speculation effectively bypasses the \code{UMIP} feature, which is supposed to prevent user-mode execution of a set of instructions, including \code{SMSW}.
Similar to the previous case, this leak contradicts AMD's claims that their CPUs are not vulnerable to RSRR.

Since the leaked information is not considered security sensitive, AMD is not planning to fix this effect.

\myparagraph{Zen1 R- and W-bit violations}
We also identified violations on AMD1 in R-bit and W-bit configurations. However, AMD was unable to reproduce the violation when testing a CPU with an equivalent microarchitecture, microcode, BIOS, and operating system. We thus concluded that a more detailed investigation of this violation is not feasible at the moment.

\subsubsection{Corroborated Findings}
\label{subsec:corroborated}

Beyond the new findings, we also detected a number of previously-known leaks:
\begin{itemize}
    \item Testing campaigns with \code{MEM A/D-bit} and \code{MEM EPT A/D-bit} configurations produced violations that are caused by \emph{MDS}~\cite{Fallout,ZombieLoad} and its variants.
    \item Campaigns with \code{MEM P-bit} configuration produced violations that are caused by \emph{Foreshadow}~\cite{weisse2018foreshadow} and its read-modify-write variant~\cite{Hofmann23}.
    \item Campaigns with \code{MEM U-bit} configuration produced violations that are caused by \emph{Meltdown}~\cite{Lipp2018}.
    \item Campaigns with \code{COMP DSS} configuration produced violations that are caused by \emph{DSS}~\cite{Hofmann23}.
    \item Campaigns with \code{REG FSGS} configuration produced violations that are caused by \emph{Rogue System Register Read}~\cite{RSRR}.
          As we tested the machines with the most recent microcode, we have not detected the original versions of this leak, but we managed to find instances of the variant reported in ``Reviving Meltdown 3a''~\cite{weber2023reviving}.
\end{itemize}

Notably, we found violations in all experiments where we expected to find one.

\subsubsection{Testing Microarchitectural Isolation Primitives}
\label{subsec:sw-validation}

To showcase the effectiveness of our tool for testing microarchitectural primitives, we re-run the tests that detected known leaks, but this time with the corresponding patches applied (if available).
We applied patches by modifying the templates: we added the corresponding instruction sequences (e.g., \code{VERW}) before transitioning to the actor representing an attacker.
We managed to find all issues in the patches that we expected to find, and we confirmed that the patches are effective in preventing the leaks.

\myparagraph{MDS Patches} Intel recommends~\cite{IntelMDS} two alternative patches: \code{VERW} and \code{L1D\_FLUSH\_CMD}.
We found that both patches are effective in preventing MDS-caused leakage.
We also tested \code{WBINVD} instruction as it has similar behavior to cache flush (although it is \emph{not} recommend by Intel);
it did not prevent MDS-caused leakage.

\myparagraph{Foreshadow Patches} Intel recommends~\cite{IntelL1TF} \code{L1D\_FLUSH\_CMD} instruction as a mitigation, which our tests confirmed to be effective.
Similarly to MDS, we tested \code{WBINVD} and found it to be ineffective as a mitigation.

\myparagraph{DSS Patches} Linux implements a patch~\cite{LinuxDSS2} that executes a dummy division of 1 by 1 before transitioning to user mode.
We tested this patch and found it to be effective.
However, before this patch was introduced, there was another patch~\cite{LinuxDSS1} that put the division in the exception handler instead of the system enter routine.
Notably, this patch was close to being merged into the mainline kernel, and it was replaced with the current patch only after a review by the security community~\cite{PhoronixDSS}.
We tested this patch with our tool, and it detected that the patch is ineffective.

\myparagraph{Meltdown Patches} A common software mitigation is to enable Kernel Page Table Isolation (KPTI)~\cite{LinuxKPTI}.
We emulate this by modifying the page table entries corresponding to the kernel actor before switching to the user actor.
We implement three variants:
\begin{inparaenum}[(i)]
    \item labeling the kernel pages as invalid (i.e., clearing the Present bit); this mitigation did not prevent Meltdown
    \item completely unmapping the kernel pages (i.e., zeroing the page table entry); this mitigation prevented Meltdown but we still found violations where user-to-kernel accesses caused MDS
    \item combining the unmapping with the \code{VERW} instruction; we found no violation with this mitigation.
\end{inparaenum}
\subsection{Performance and Detection Time}

The performance of our tool varies across the testing targets and configurations, but most violations are detected within less than four hours and all tests complete within 24 hours. In more details,
\begin{asparaitem}
    \item The tool performed hardware measurements at the rate ranging from 800 to 4500 measurements (i.e., hardware traces) per second.
    The performance depended on the type of template (e.g., VMs are slower than user processes), the number of actors, and the complexity of the operations performed by the actors.
    \item In terms of overall testing throughput, the tool ran all test cases through completion (i.e., 100k test cases each with 50 inputs) within 2--24 hours for \code{MEM} configurations, within 3-8 hours for \code{COMP} configurations, and within 1.5-11 hours for \code{REG} configurations.
    \item We measured the average detection time for each violation by re-running the experiments with violations 10 times with different seeds.
    The tool detected 85\% of the violations within one hour, and 98\% of the violations within four hours.
    In terms of the number of test cases, the tool detected 85\% of the violations within 20K rounds, and 93\% of the violations within 40K rounds.
    The tool failed to detect a violation in 4\% of these tests.
\end{asparaitem}

These numbers demonstrate that our design choices (e.g., custom VM management~\secref{executor}, adaptive sample size~\secref{noise}) yield a practical tool for testing microarchitectural isolation.

\section{Discussion}
\label{sec:discussion}

This section discusses some aspects of our work.

\myparagraph{Kernel Module vs Custom OS}
In our implementation, we chose to implement the measurement environment within a kernel module.
An alternative approach would be to implement a completely custom operating system that would take care of setting up the measurement environment.

The advantage of a custom OS would be that we would not have to deal with preservation of the host OS state, and it could potentially reduce the level of noise in the measurements.
However, the complexity of using a custom OS would be significantly higher compared to using a Linux kernel module.
The implementation as a Linux kernel module that we provide will maximize adoptability and maintainability in practice.

\myparagraph{Modeling Choices}
For our purpose, it was sufficient to just collect contract traces for the victim execution. Our modeling framework can be easily generalized to support multiple actors with actor-specific contracts, whose traces are concatenated into a single contract trace.
For contracts that model speculative execution \cite{Guarnieri2021,Hofmann23},
we must then ensure to not speculatively transition between different actors (e.g., when a VMEXIT lies on a branch that is not taken architecturally).
We also decided to treat VMX instructions as simple jumps in the model.
Theoretically, this does not have to be the case, and we could assign more involved semantics that, e.g., reveal the ID of the loaded guest.

\myparagraph{Limitations}
The search space for fuzzing the microarchitectural isolation is vast.
We implemented the essential components, but there are several aspects that our tool does not yet cover.
First, architectural information flows between actors, e.g., shared memory, are not supported.
Second, self-modifying code is not supported.
Finally, the location of control structures (e.g., IDT, GDT) is fixed, and hence, if a CPU makes predictions on the location of control structures, we would not be able to detect corresponding leaks.
These features would make the tests more complete, but we considered them a lower priority than the experiments presented in this paper.
We consider implementing these features as future work.

Another limitation is the impact of the measurement framework on the results.
Our tool inserts certain code sequences into the template (e.g., macros to collect hardware traces), and these sequences necessarily affect the microarchitectural state.
These changes could potentially affect the results and, in the worst case, prevent the tool from detecting a leak.
However, this effect is compensated by running large testing campaigns with many test cases, and we also intentionally build the tool to minimize the interference (e.g., by reserving a subset of registers to avoid memory accesses).

Beyond the limitations of the tool, our evaluation tests only a subset of possible leaks between security domains, and for example, does not cover cross-domain branch training or other control-flow attacks.
Moreover, some of the leaks detected in our evaluation required a large number of measurements, which suggests that the duration of our experiments may not be sufficient to detect all possible leaks.
Doing a complete evaluation would be a significant undertaking, beyond the scope of this paper.
But, given the findings from our evaluation, we believe that a dedicated and comprehensive evaluation study would be valuable, and we consider it as future work.

\myparagraph{Application as a Prototyping Tool}
This paper focused on one use-case: automated testing of microarchitectural isolation primitives.
However, the proposed tools can also be used for prototyping and sharing prototypes of speculative execution attacks.
Our tool allows its user to write the test case completely manually (in contrast to random-generation mode used in this paper).
Accordingly, when a security researcher discovers a new speculative execution attack, they can use the tool to build a PoC of the attack, and then share the PoC with the CPU vendor.
This provides several benefits compare to the current practice of writing stand-alone PoCs:
(1) the tool takes care of the low-level system management tasks (e.g., configuring the CPU and performing cross-security domain transitions), and it comes with built-in measurement capabilities, so the PoC can be built quickly and easily (e.g., Foreshadow could be demonstrated with less than 30 lines of template assembly);
(2) the tool provides a stable and low-noise measurement environment, which makes it easier for CPU vendors, as well as other affected parties, to reproduce the attack and to assess its impact.
As a result, the tool can help to streamline the process of reporting and fixing speculative execution attacks.

\myparagraph{Porting to Other Targets}
While our current implementation targets x86 CPUs and VM/User/Kernel isolation, the framework is designed to be easily portable to other platforms and isolation primitives.
The main challenge in porting the tool to other platforms is to implement the corresponding execution environments.
To test other isolation abstractions (e.g., memory protection techniques like Intel MPK), we would need to modify the executor to enable the corresponding isolation primitive and to configure the corresponding page table bits.
To port the tool to other architectures (e.g., ARM, RISC-V), we would need an executor that implements system management functionality for the corresponding architecture, such as creating VMs/user processes, setting up their address spaces, etc.
In addition, other architectures will require the generator to handle new instruction types.

\myparagraph{Testing for Injection-type Leaks}
In the evaluation, we focused on the leaks that extract information from a victim.
However, an inverse type of the attack is also possible, where the attacker microarchitecturally injects information into the victim to manipulate its execution, as in LVI~\cite{VanBulck2020} and Spectre V2~\cite{Kocher2018} attacks.

Our tool can be used to detect these leaks via an inverse template: 
The execution starts from the attacker's (random) code, which injects attacker's data into the microarchitectural state.
Then, the template switches to the victim's code, which executes its own (random) code.
The victim's execution is monitored by collecting its hardware traces.
If the victim's hardware traces change depending on the attacker's data, this implies that the attacker can control the victim's execution, and it will be reported as a violation.

\section{Related Work}
\label{sec:related}

\myparagraph{Automated Black-Box CPU Testing}
There are several tools that automatically test black-box CPUs for microarchitectural vulnerabilities, but all of them either run exclusively in a single trust domain or test for variants of known vulnerabilities.
None of the related approaches can automatically detect novel vulnerabilities across domain boundaries.

Revizor is a model-based relational testing (MRT) tool that tests microarchitectural leakage of a CPU against formal leakage contracts~\cite{oleksenko2022revizor,oleksenko2023hide,Hofmann23}, as discussed in~\subsecref{mrt}.
Revizor runs only in kernel space.

Scam-V~\cite{Nemati2020a,buiras2021micro} is another MRT tool, which, unlike Revizor, uses symbolic execution to automatically generate inputs that yield the same modelled leakage. 
Similarly to Revizor, this tool can detect unknown leaks, but only within a single trust domain (ARM TrustZone in this case).

SpeechMiner~\cite{Xiao2020} examines the combination of known Meltdown-type gadgets and measures their exportability quantitatively in terms of the underlying race conditions.
This tool also works in a single domain (user or kernel), and cannot test transitions between them.

Transynther~\cite{Moghimi2020a} is a tool that randomly combines building blocks of known Meltdown and MDS-type vulnerabilities to do variant search.
Transynther runs two sibling threads, one victim that fills microarchitectural buffers with known values and one attacker that runs the code snippet under test.
This tool differs from our work in that it focuses on a specific type of leakage: Meltdown and MDS.
As such, it cannot find entirely new types of leakage.

RegCheck~\cite{weber2023reviving} specializes on Meltdown-3a, a vulnerability that leaks the contents of privileged registers into user space. The tool automatically checks which registers are susceptible to the attack on various CPUs. Similarly to Transynther, RegCheck is specialized on a specific type of leakage and cannot find entirely new types of leakage.

\myparagraph{Automated Side Channel Detection}
Other approaches expose new (variants of) timing side channels.
Osiris~\cite{Weber2021} observes timing differences in randomly generated code snippets.
Plumber~\cite{0002NSTR22} generates abstract code templates and input conditions from code that is known to trigger distinguishable behavior.
While the leaks detected by these tools could potentially cross the domain boundaries, they do not explicitly test for this.

\myparagraph{Automated White-Box CPU Testing}
White-box tools analyze pre-silicon CPU designs for microarchitectural vulnerabilities. 
These tools need access to a description of the microarchitecture and are therefore not applicable to commercial CPUs.
Examples are IntroSpectre~\cite{Ghaniyoun2021}, which analyzes an RTL description, and CheckMate~\cite{Trippel2018}, which builds on memory consistency models.

\myparagraph{Validating Microarchitectural Isolation Primitives}
Past findings have demonstrated the need for a thorough analysis of isolation primitives, as both hardware and software mitigations have been compromised (e.g. Enhanced and Automatic IBRS~\cite{bhi, inception}, Fine Indirect Branch Tracking~\cite{inspectreGadget}, retpoline~\cite{wikner2022retbleed}, and LFENCE/JMP~\cite{milburn2022win} against variants of Spectre v2 attacks, buffer overwriting (via \texttt{VERW}) against MDS~\cite{cacheout}); for a recent overview see~\cite{quarantine}.
Microarchitectural isolation primitives are hard to validate, though.
\footnote{
Russell Currey discusses the difficulties of validating kernel patches in this insightful talk: \url {https://www.youtube.com/watch?v=GhNMbXQci28}.
}
Traditionally, mitigations are analyzed either by testing against a PoC or by purely theoretic reasoning~\cite{Canella2020}.
An exception are obfuscating side-channel defense schemes like randomized mapped caches and access timing randomization.
Due to the underlying randomization, these defense schemes leave some residual leakage, which can be quantified~\cite{metior,casa}.
Our approach that randomizes the code executed by the victim and the attacker thoroughly tests if mitigations prevent information leakage across trust domains. Like this, our tool would even be able to find novel vulnerabilities introduced by a mitigation.

\section{Conclusion}
\label{sec:conclusion}

We introduced a new fuzzing approach based on model-based relational testing to test microarchitectural isolation between security domains.
Using this method, we systematically tested the microarchitectural isolation between security domains on a set of recent x86 CPUs and the robustness of contemporary software defenses against microarchitectural leakage.
To achieve this, we introduced the actor abstraction to represent adversary security domains.
Based on actors, we built template-based test cases and a dedicated execution environment for running them.
To circumvent opaque microcode executed upon some transitions, we introduced a new statistical measurement technique.
This new testing approach led to the discovery of a range of new cross-domain leaks.

\section*{Acknowledgments}

We would like to thank Adrien Ghosn and Kaveh Razavi for the discussions and support they provided.
We also thank our shepherd and the anonymous reviewers for their valuable feedback.
We are grateful to AMD and Intel PSIRT teams as well as Microsoft Security Response Center for their prompt responses and the constructive discussions we had while reporting the vulnerabilities.

This work was supported in part by a Microsoft Swiss JRC grant and by the Swiss State Secretariat for Education, Research and Innovation under contract number
MB22.00057 (ERC-StG PROMISE).

\bibliographystyle{IEEEtran}
\bibliography{ms.bib}
\balance

\appendices 
\label{appendix}
\newpage

\section{Instruction Set Tested in Evaluation} 
\label{appendix:isa}

\subsection{Baseline Instruction Pool}
\label{app:isa-base}
Our experiments use instructions from the following pool:
\texttt{
    (LOCK) ADD/ADC, (LOCK) SUB/SBB, (LOCK) INC, (LOCK) DEC, (LOCK) NEG, (I)MUL, (I)DIV, (LOCK) AND, (LOCK) NOT, (LOCK) OR, (LOCK) XOR,
    BSF, BSR, BT, (LOCK) BTC, (LOCK) BTR, (LOCK) BTS, BSWAP, 
    MOV, MOVSX, XCHG, CMP, CBW, CDQ, CWD, CWDE, CLC, CLD, CMC, LAHF, SAHF, STC, STD, 
    CMOVB, CMOVBE, CMOVL, CMOVLE, CMOVNB, CMOVNBE, CMOVNL, CMOVNLE, CMOVNO, CMOVNP, CMOVNS, CMOVNZ, CMOVO, CMOVP, CMOVS, CMOVZ, (LOCK) CMPXCHG, (LOCK) XADD, LEA, NOP, SET*
}

\subsection{Instructions in System Register Read}
\label{app:isa-extended}
The experiments testing for leaks when reading system registers add instructions and registers to the instruction pool, one sub-class at a time as follows:
\begin{asparaitem}
    \item \textbf{Register:} \texttt{RDTSC(P), CR[0,2,3,4,8], DR[0,6,7]}
    \item \textbf{MSR:} \texttt{RDMSR, WRMSR}
    \item \textbf{XSAVE:} \texttt{XGETBV, XSETBV} (Intel, AMD) and \texttt{XRSTORS, XSAVES} (Intel)
    \item \textbf{UMIP:} \texttt{LGDT, LLDT, LIDT, LTR, SGDT, SIDT, SLDT, STR, SMSW, LMSW}
    \item \textbf{FSGS:} \texttt{RDFSBASE, RDGSBASE, WRFSBASE, WRGSBASE, SWAPGS}
    \item \textbf{VM (Intel):} \texttt{VMCLEAR, VMLAUNCH, VMPTRLD, VMPTRST, VMRESUME, VMXOFF, VMXON, INVVPID}
    \item \textbf{VM (AMD):} \texttt{VMMCALL, VMLOAD, VMRUN, VMSAVE}                    
    \item \textbf{SEC (Intel):} \texttt{GETSEC, PCONFIG, ENCLS, ENCLV}
    \item \textbf{SEC (AMD):} \texttt{PSMASH, PVALIDATE, RMPADJUST, RMPQUERY, RMPUPDATE, SKINIT}                    
    \item \textbf{MISC:} \texttt{CLAC, STAC, CLGI, STGI, CLTS, HLT, INVD, INVLPG, INVPCID, SYSRETQ, SYSEXITQ, WBINVD, WBNOINVD} (Intel, AMD) and \texttt{TLBSYNC} (AMD)    
\end{asparaitem}

\end{document}